\documentclass{aa}  

\usepackage{graphicx}
\usepackage{natbib}
\usepackage{subfigure}
\usepackage{bm}
\usepackage{orcidlink}
\usepackage{hyperref}
 
\usepackage{subcaption}
\usepackage{txfonts}
\usepackage{comment}

\usepackage[switch]{lineno}

\begin{document}

\title{Effect of gravity darkening and oblate factor in rapidly rotating massive stars}

\author{Bhawna Mukhija \orcidlink{0009-0007-1450-6490}
          \inst{1}
          \and
        Michel Curé \orcidlink{0000-0002-2191-8692} \inst{2}
        \and
        Ignacio Araya  \orcidlink{0000-0002-8717-7858}  \inst{3}
          \and
        Catalina Arcos  \orcidlink{0000-0002-4825-4910} \inst{2}
        \and
        Alejandra Christen \orcidlink{0000-0003-0763-7912} \inst{4}
         }

   \institute{Ariel University, Ariel, 4070000, Israel\\
              \email{bhawnam@ariel.ac.il}
         \and
             Instituto de F\'{\i}sica y Astronom\'{\i}a, Facultad de Ciencias, Universidad de Valpara\'{\i}so,
             Av. Gran Breta\~na 1111, Valpara\'{\i}so, Chile
         \and     
              Centro Multidisciplinario de F\'isica, Vicerrector\'ia de Investigaci\'on, Universidad Mayor, 8580745 Santiago, Chile
         \and
             Instituto de Estad\'{\i}stica, Facultad de Ciencias, Universidad de Valpara\'{\i}so,
             Av. Gran Breta\~na 1111, Valpara\'{\i}so, Chile
                     } 

   \date{}

  \abstract
  {Rapid rotation in massive stars leads to gravity darkening and oblateness, significantly affecting their radiation-driven winds. These effects can alter wind dynamics and play a role in forming slowly equatorial outflowing winds.}
   {This work investigates the transition region where the fast solution (i.e. high terminal velocities) of radiation-driven winds in a massive rotating star, in the frame of the modified-CAK theory, switches to the $\Omega$-slow solutions (a denser and slower wind) when the effects of gravity darkening and oblateness are considered. This $\Omega$-slow solution appears when the rotational speed is  $\gtrsim 75\%$  of the critical rotation speed.}
   {To explore the transition region for various equatorial models of B-type stars, we focus on the co-existence interval where both solutions simultaneously exist and the transition point where fast solutions switch to $\Omega$-slow solutions. }
   {Using our stationary numerical code {\sc{Hydwind}}, we first analyse the individual effects of gravity darkening and stellar oblateness caused by high rotational speeds and then examine their combined impact on the wind solutions.}
   {We find that for a certain range of rotational speeds, both the fast and Ω-slow solutions can co-exist, and the co-existence range strongly depends on the initial conditions. When only gravity darkening is considered, the co-existing interval shifts towards higher rotational speeds. While in the presence of the oblateness, the co-existing interval also occurs at higher rotational speeds; however, it is less than the gravity darkening effect. We also explored how line-force parameters affect the critical point, the location of the co-existing interval, and where the solution switches.
   }

\maketitle

\section{Introduction}

Massive stars play an important role in understanding the formation and structure of the universe \citep[][]{2012ASSL..384.....D, 2012ARA&A..50..107L, 2015HiA....16...51V, 2016ApJ...817...66K,2022MNRAS.514.3736S,2022MNRAS.512.4116F, 2024ApJ...974..124M, 2025ApJ...986..188M}. 
Due to their enormous masses and powerful stellar winds, these stars significantly contribute to the chemical enrichment of the universe by synthesising heavy elements \citep[][]{2012ARA&A..50..107L,2013ascl.soft03015H, Nomoto2013, 2019Sci...363..474J, 2024ApJ...961...89V}. The mechanism driving these powerful winds was first described by \citet{1975ApJ...195..157C} and referred to as the  CAK-theory of radiation-driven winds.  According to this theory, the radiative acceleration in the atmosphere arises from a multitude of spectral absorption lines, leading to the outward expulsion of material \citep[][]{2000ARA&A..38..613K, 2008A&ARv..16..209P}. Later on \citet[][]{1986ApJ...311..701F, 1986A&A...164...86P} introduced the modified-CAK (m-CAK) model, which incorporates two key enhancements: it accounts for the finite size of the star through the finite disk correction factor, and it includes the effects of stellar rotation, specifically the centrifugal force resulting from rapid rotation. This m-CAK model has successfully described both terminal wind velocities ($V_{\rm \infty}$) and mass-loss rates ($\dot M$) observed in massive stars \citep[][]{2008A&ARv..16..209P}. They found that stellar rotation has a significant impact on the observable wind parameters of massive stars. However, they could not find wind solutions for stars rotating at high rotational speeds (i.e. when $v_{\rm{rot}}$\footnote{As a reference, a rotational speed $v_{\rm{rot}} \sim 500 - 550 \,$ km/s is equivalent to $\Omega \sim 90 - 99\%$.} is more than 75$\%$ of its critical velocity, $v_{crit}$). Stars with such rotation rates (e.g. Oe, Be, Bn, WR, among others) are categorised as rapid rotators, as their high rotational speed significantly influences their wind dynamics and evolution \citep[][]{1991A&A...244L...5L, 1993ApJ...409..429B, 1996ApJ...472L.115O, 2000ARA&A..38..143M, 2004MNRAS.350..189T,2023Galax..11...68C}. 

\citet{2004ApJ...614..929C} revisited the effect of stellar rotation on 
radiation-driven winds using 1D stationary models and identified a novel wind 
solution that emerges when the rotational speed exceeds $\sim 75\% $ of the 
critical limit. This solution, referred to as the $\Omega$-slow solution 
(here $\Omega$ is a dimensionless rotation parameter), is defined as the ratio 
of the stellar angular velocity to the critical angular velocity,
\begin{equation}
\Omega = \frac{\omega_{\rm rot}}{\omega_{\rm crit}} 
= \frac{v_{\rm rot}(\Omega,\mathrm{eq})}{v_{\rm crit}}
\frac{R_{\rm eq}^{\rm max}}{R(\Omega,\mathrm{eq})},
\end{equation}
where $v_{\rm rot}(\Omega,\mathrm{eq})$ is the equatorial rotational velocity, 
$R(\Omega,\mathrm{eq})$ is the equatorial radius for a given $\Omega$, and 
$R_{\rm eq}^{\rm max}=3R_{\rm pole}/2$ is the maximum equatorial radius at critical 
rotation. A slower and denser wind characterises the $\Omega$-slow solution
compared to the standard m-CAK solution (henceforth referred to as the fast solution). 
It predicts significantly enhanced densities near the equator, particularly when 
combined with bistability effects \cite{1991A&A...244L...5L}, see also 
\citet{2021A&A...648A..36B,2023A&A...676A.109B,2024A&A...687L..16D}, resulting in a 
density contrast of up to $10^{4}$ between the equatorial and polar regions. 
This suggests that the $\Omega$-slow solution could play a critical role in the 
formation of equatorial disks observed in B[e] stars \citep[][]{2005A&A...437..929C}. 
\citet{2007ApJ...660..687M} conducted detailed time-dependent hydrodynamic simulations 
to investigate the behaviour of radiation-driven winds in rapidly rotating stars. 
Their findings suggested the inviability of the $\Omega$-slow solution as a mechanism 
to reproduce the high equatorial densities inferred for B[e] star disks. They also 
revealed the presence of abrupt kink transitions in the wind velocity profiles at 
rotational speeds ranging from 75\% to 85\% of the critical rotational velocity.

\citet{2018MNRAS.477..755A} investigated the transition region in a 1D equatorial model of a rapidly rotating star, focusing on where the fast radiation-driven wind solution switches to the $\Omega$-slow solution at high rotational speeds. They found a co-existing region inside a small interval of $\Omega$, where both solutions are simultaneously present at the equatorial plane, and satisfy the same boundary conditions at the stellar surface. Their calculations were based on the stationary code {\sc{Hydwind}} \citep{2004ApJ...614..929C} and time-dependent hydrodynamical code {\sc{Zeus-3D}} \citep{1996ApJ...457..291C}. They suggested that the topology of the non-linear m-CAK differential equation predicts two wind solutions as a function of $\Omega$ for a given set of line force parameters. The fast solution ceases to exist when $\Omega$ is above a certain threshold value that depends on the stellar and line force parameters.
In contrast, the $\Omega$-slow solution begins to exist from this interval up to $\Omega\,$=\,1. They also showed that time-dependent solutions are highly sensitive to the topology branch to which the initial velocity profile belongs. If the initial velocity profile deviates significantly from the steady-state asymptotic solution, the calculations can result in non-monotonic 'kink' solutions. To achieve globally monotonic acceleration and obtain either the fast or $\Omega$-slow solutions, an initial velocity profile must accurately represent the respective fast or slow regime (topology branch). They also revealed that higher velocities and lower densities characterise the fast solution regime, whereas the $\Omega$-slow solution regime showed lower velocities and higher densities. They also found that stellar and line-force parameters influenced the position and width of the co-existence region. An increase in the value of the line-force parameter $\alpha$ shifts the co-existence region to higher $\Omega$ values, while the width of the region remains unchanged. Additionally, they did not account for gravity-darkening effects and the distortion of the star's shape due to high rotational speeds; both have significant implications for the structure and dynamics of the wind.

When accounting for high rotational velocities, a star becomes oblate, adopting a shape resembling a rotating ellipsoid. According to \citet{1924MNRAS..84..665V}, the radiative flux F at a given co-latitude on a rotating star is directly proportional to the local effective gravity, i.e. $ T_{\rm eff} \propto {g_{\rm eff}}^{1/4}$. The oblateness caused by rapid rotation modifies the local effective gravity, which is expressed as $\vec{g}_{\rm eff} = \vec{g}_{\rm grav} + \vec{g}_{\rm rot}$. Consequently, this variation in $g_{\rm eff}$ alters the surface temperature distribution of the star, with polar regions becoming hotter and equatorial regions cooler, a phenomenon referred to as Gravity Darkening \cite[hereafter GD, see][]{1924MNRAS..84..665V}.  Recent interferometric observations of nearby rapidly rotating stars by \citet[][]{2005AAS...207.8204B, 2006ApJ...651..617A, 2006ApJ...637..494V, 2010RMxAC..38..117Z} indicate that GD does not conform well to von Zeipel’s model, which appears to overestimate the temperature difference between the poles and the equator. This discrepancy is often expressed with a power law, $T_{\rm eff} \propto g_{\rm eff}^\beta$, where the exponent $\beta$ is less than 1/4. The reduced variation in temperature with latitude, compared to von Zeipel’s model, is traditionally attributed to a thin convective layer at the star's surface, as described by \citet{1967ZA.....65...89L}, who found $\beta = 0.8$ for stars with convective envelopes. Similar observations were made by \citet[][]{2006ApJ...643..460L, 2007A&A...470.1013E, 2011A&A...533A..43E,Gagnier2019} in their 2D numerical models of rotating stars.

In this study, we extend the work of \citet{2018MNRAS.477..755A} by examining the co-existence region where the fast solution transits to the $\Omega$-slow solution, in the presence of the GD and oblateness. Our analysis focuses on a 1D equatorial model with high rotational speeds. Initially, we recalculated the co-existing interval, consistent with the findings of \citet{2018MNRAS.477..755A}. We then explored wind solutions while accounting for the influence of GD and oblateness effects. Furthermore, we also explored the dependence of the co-existing region on various stellar and line-force parameters.

This paper is organised as follows: Section \ref{sec2} introduces
our hydrodynamic wind model, starting with our assumptions. Then, we outline the time-dependent and steady-state equations used to describe the rotating m-CAK theory. In Section \ref{sec3}, we solve the stationary equations in the star’s
equatorial plane for a wide range of rotational speeds where
fast and $\Omega$-slow solutions are achieved in the presence of GD and oblateness. In addition, a co-existence region is found where both solutions are simultaneously
present. The influence of the line-force parameters on the co-existing regions is also studied. Finally, in Section \ref{sec4}, we present the discussion and summary of our findings.

\section{1D steady state wind solutions for fast rotating stars}
\label{sec2}
The work of \citet{2017ApJ...846....2A} thoroughly developed the complete set of equations governing the m-CAK radiation-driven wind theory, including the oblate finite disk correction factor and the gravity darkening effect models from \citet{1924MNRAS..84..665V} and \citet{2011A&A...533A..43E}, and incorporating these effects into the hydrodynamic wind model implemented in the {\sc{Hydwind}} code. It is worth noting that, for the m-CAK theory, we have also implemented a time-dependent version of the Equation of Motion (EoM) using the {\sc ZEUS-3D} code \citep[][and references therein]{1996ApJ...457..291C,2010ApJS..187..119C}. Although these two numerical methods differ significantly ({\sc{Hydwind}} solves the stationary m-CAK EoM, and {\sc ZEUS-3D} handles its time-dependent version), the steady-state solutions they produce are in close agreement. Given this consistency, and considering that {\sc{Hydwind}} is explicitly developed for the m-CAK theory, we decided to implement gravity darkening and stellar oblateness effects within this code. While {\sc ZEUS-3D} is a well-established, versatile astrophysical (Magneto-)hydrodynamic code widely used for various applications, {\sc{Hydwind}}'s specialisation makes it our preferred tool for m-CAK stationary wind modelling.\\

The reader is referred to \citet{2017ApJ...846....2A} for the comprehensive set of equations described therein, as this work lays the essential groundwork and provides the mathematical formalism necessary to understand the advancements presented here.
The primary goal of this work is to investigate and delineate the regions of co-existence of wind solutions subject to the combined effects of gravity darkening and stellar oblateness, thereby advancing our understanding of rotationally modified radiatively driven stellar winds.
All solutions presented in this work correspond to the equatorial plane (co-latitude $\theta=90^\circ$).

\subsection{Radiation-driven wind equations}

The 1D m-CAK hydrodynamical equations for rotating line-driven 
winds are formulated from the continuity and momentum equations, with the 
radiative acceleration parametrised following the standard CAK formalism 
\citep{1975ApJ...195..157C,1986ApJ...311..701F}. The basic equations are:
\begin{equation}
\label{continuity}
4 \pi r^{2}\, \rho \, v = \dot{M}\, ,
\end{equation}
and
\begin{equation}
\label{momentum}
v \, \frac{dv}{dr}=-\frac{1}{\rho}\frac{dP}{dr} - \frac{G\, M \, (1-\Gamma_{\mathrm{E}})}{r^{2}} 
+ \frac{v_{\phi}^{2}(r)}{r} + g_{\mathrm{rad}}^{\mathrm{line}}(\rho, dv/dr, n_{\mathrm{E}})\, ,
\end{equation}
where $r$ is the radial coordinate, $M$ the stellar mass, $\rho$ the density, $v$ the radial velocity, 
$P$ is the gas pressure for an isothermal wind, and $\Gamma_{\mathrm{E}}$ the Eddington parameter due to electron scattering.  
For an ideal gas, the equation of state is given by $P=a^{2}\rho$, being $a$ the isothermal sound speed.

Following \citet{2004ApJ...614..929C} and \citet{2017ApJ...846....2A}, 
we adopt the usual variable transformations $u=-R_{*}/r$, 
$w=v/a$, and $w'=dw/du$, with $a_{\mathrm{rot}}=v_{\mathrm{rot}}/a$, 
to write the compact form of the EoM as
\begin{eqnarray}
\nonumber
\label{motion-eq}
F(u,w,w') & = & \left( 1- \frac{1}{w^{2}} \right) w \frac{dw}{du} + A + \frac{2}{u} + a^{2}_{\mathrm{rot}}\, u \\
& - & \, C' \, f_{\rm OFD}(\Omega,\theta,u)\, g(u)\, w^{-\delta} 
\left( w \frac{dw}{du} \right)^{\alpha}  = 0\, ,
\end{eqnarray}
where $f_{\rm OFD}$ is the oblate finite-disk correction factor including 
gravity darkening. Details of the derivation and definitions of the parameters are given in \citet{2004ApJ...614..929C} 
and \citet{2017ApJ...846....2A}. \\
Here, $\alpha$, $k$, and $\delta$ are the standard m-CAK line-force parameters 
\citep{1982ApJ...259..282A, 2000A&AS..141...23P}. 
The parameter $\alpha$ describes the ratio between the line force from optically thick lines and the total line force. 
The parameter $k$ (absorbed into the eigenvalue $C'$ in Eq.~\ref{motion-eq}) is
related to the number of lines contributing effectively to the driving of the wind. 
Finally, $\delta$ accounts for changes in ionisation throughout the wind.

To solve the m-CAK equation of motion (Eq.~\ref{motion-eq}) while considering GD and the oblate distortion of a rapidly rotating star, we used the simplified approximation from \citet{2017ApJ...846....2A} for the $f_{\rm OFD}$ that makes it possible to solve the EoM in {\sc{Hydwind}}. This approximation preserves the known topology, and \citet{2017ApJ...846....2A} solved it in both equatorial and polar directions.

\subsection{Model setup}
\begin{figure}
    \centering
    \includegraphics[width=\linewidth]{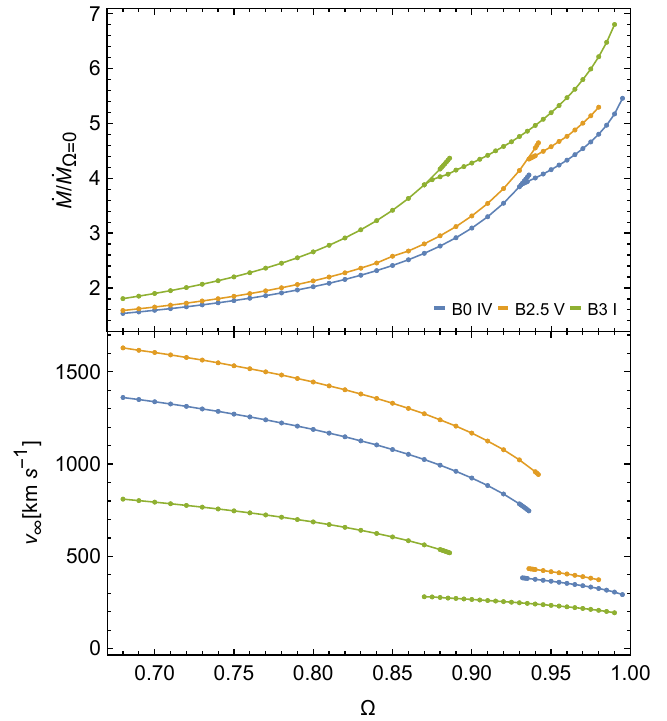}
    \caption{Mass-loss rate and terminal velocity as a function of the rotational speed.
    The upper panel shows the mass-loss rate, normalised to the non-rotating case. The lower panel shows the terminal velocities. These values are obtained using a stationary hydrodynamic solution with the {\sc{Hydwind}} code, accounting for the gravity darkening (GD) effect.}
    \label{1}
\end{figure}

Following \citet{2017ApJ...846....2A}, we used the hydrodynamic code {\sc{Hydwind}} \citep{2004ApJ...614..929C} to solve the EoM.
This code uses a $\beta$-law as initial trial solution, with $\beta$ values in the range  $\approx 0.8 - 1.2$ for fast solutions and 
$\approx 3.0 - 3.5$ for $\Omega$-slow solutions \citep{2018MNRAS.477..755A}, and terminal velocity speeds of 1000 km/s and 400 km/s, respectively. 
We begin our analysis by calculating the wind solution in the presence of the GD for the interval  $0.68 < \Omega<0.99$. Our calculations are typically carried out for main-sequence B-type stars, specifically B0 IV, B2.5 V, and B3 I. For the B2.5V star, we follow \citet{2007ApJ...660..687M}, with stellar parameters set as: $T_{\rm eff}$=$20 \,000 $ K, $R_{\ast}=4~R_{\odot}$, and $\log g$ = 4.11 dex. See  Table \ref{1}, last row, for the base density value. The line force is parametrised following \citet{1995ApJ...454..410G}, with the values  $\alpha = 0.5$, $\delta=0.0$, and $k=0.6098$ or line strength parameter $\Bar{Q}=1533$. For the B0 IV star, we use the parameters from \citet{2007ApJ...668..481S}, while the line force parameters are taken from \citet{1986A&A...164...86P}.  The stellar and line-force parameters for the B3 I model are adopted from the equatorial bi-stable wind model developed by \citet{2000A&A...359..695P}. The stellar and line force parameters of these three models are summarised in Table~\ref{T1}.

The assumption of invariant m-CAK parameters ($\alpha$, $k$, and $\delta$) with varying stellar rotation ($\Omega$) is an inherent approximation. These parameters are fundamentally derived from atomic line transitions, which are sensitive to local physical conditions that can be altered by stellar rotation through oblateness and gravity darkening. There are a few works that address the calculation of the line-force parameters for the fast solutions \citep{1986A&A...164...86P,2000A&AS..141...23P,2015MNRAS.453.3120N,2019ApJ...873..131G}, but no studies have performed these calculations for the $\Omega$-slow solution. In this work, following \citet{1994A&A...292..221S} and \citet{1995A&A...298..179D}, we vary the value of the line force parameter $\alpha$, which represents the ratio between the line-force originating from optically thick lines and the total line-force that controls the contribution of optically thick and thin lines.

\begin{table}
\caption{Stellar and line-force parameters for our models.}
    \centering
    \begin{tabular}{l c c c c c c c c c}
    \hline 
    \hline
Model  & B0 IV & B2.5 V & B3 I\\
\hline
   $T_{\rm eff}$ [kK] & 25.0 &20.0& 17.5 \\ 
    log g & 3.50 & 4.11 & 2.70 \\
    $R_{*} [R_{\odot}]$ & 10.00 & 4.00 &  47.00 \\ 
    $\alpha$ & 0.565 & 0.500 & 0.450 \\
    $\delta$ & 0.02 & 0.00 & 0.00 \\
    k & 0.32 & 0.61 & 0.57 \\
    $\Bar{Q}$ & 2792.0 & 1533 & 361.6 \\
    $\rho_{0} ~[{\rm g~cm^{-3}}]$ & $5.0 \times 10^{-11}$ &$ 8.7 \times 10^{-13}$ & $1.0 \times 10^{-11}$\\
   \hline 
    \end{tabular}
    \label{T1}
\end{table}

\section{Co-existence of fast and $\Omega$-slow regimes}
\label{sec3}
\subsection{Numerical method for mapping solution regimes}

The existence of the two distinct solution branches (fast and $\Omega$-slow solutions) requires a robust numerical strategy to map each regime and accurately identify the interval of $\Omega$ where they may co-exist. To achieve this, we employed the following systematic procedure using the {\sc{Hydwind}} code.

\noindent Tracing the $\Omega$-slow Solution Branch:\\ 
We begin at a near-critical rotational speed (e.g. $\Omega = 0.995$), where only the $\Omega$-slow solution is expected. At this starting point, a $\beta$-velocity law with a large exponent (typically $\beta \approx 3.0 - 3.5$) is used as the initial trial solution. Once this solution converges, we gradually reduce $\Omega$ and, if necessary, fine-tune $\beta$. 
This process is continued until a value of $\Omega$ is reached, below which a converged $\Omega$-slow solution can no longer be found.

\noindent  Tracing the Fast Solution Branch:\\ Conversely, to map the fast solution branch, we start at a moderate rotational speed where the fast wind is dominant (e.g. $\Omega = 0.68$). A $\beta$-velocity law provides the initial guess with a small exponent (typically $\beta \approx 0.8$). After convergence, we incrementally increase the value of $\Omega$. This process is repeated until the fast solution no longer exists.\\

The co-existence interval is then defined as the range of $\Omega$ for which this procedure yields valid solutions for both the fast and $\Omega$-slow branches. This systematic approach ensures that each solution branch is followed continuously, providing a robust determination of the boundaries for each regime. 

\subsection{GD effect}
\begin{table*}
\caption{Properties of fast and $\Omega$-slow solutions for B-type stellar models (GD effect).}
    \centering
    \begin{tabular}{l c c c c c c c c c}
    \hline \hline 
    Model & $\Omega~\rm range$ & Solution  &  Co-existence interval& $\dot M (\Omega_{\rm i})$   & $v_{\infty}(\Omega_{\rm i})$ & $ r_{\rm crit}(\Omega_{\rm i})$ & $\dot M (\Omega_{\rm f})$ & $v_{\infty}(\Omega_{\rm f})$ &  $ r_{\rm crit}(\Omega_{\rm f})$ \\
        & &type & $\Omega_{\rm i}-\Omega_{\rm f}$ & $[\rm M_{\odot}~yr^{-1}]$ &$[ \rm km~s^{-1}]$ & $[\rm R_{\odot}]$  & [$\rm M_{\odot}~yr^{-1}]$ & $ [\rm km~s^{-1}] $& [$\rm R_{\odot}$]\\
        \hline
        B2.5 V & 0.68 - 0.942 &  Fast & 0.936 - 0.942 & $4.35 \times 10^{-9}$ & 433.4 & 21.82 & $ 4.65 \times 10^{-9}$ & 943.7 & 1.027\\
         & 0.946 - 0.995 & $\Omega$-slow & &\\ 

 \\
      B0 IV & 0.68 - 0.936 & Fast &  0.932 - 0.936 & $4.716 \times 10^{-7}$ & 383.5 & 15.64 &$ 4.908 \times 10^{-7}$ & 746.5 & 1.058   \\
      & 0.932 - 0.995 & $\Omega$-slow & \\

      \\
      B3 I & 0.69 - 0.886 & Fast & 0.870 - 0.886 & $2.552 \times 10^{-6}$ & 280.9 & 12.51 & $2.872 \times 10^{-6}$ & 519.1 & 1.049 \\

      & 0.87 - 0.995 & $\Omega$-slow & \\
      \hline
       \end{tabular}
       \label{T2}
        \tablefoot{
        The wind parameters are evaluated at the angular velocity values, \( \Omega \), corresponding to the beginning (\(i\)) and end (\(f\)) of the co-existence region. Columns 5–7 correspond to the \( \Omega \)-slow solution, while the last three columns refer to the fast solution.}
      \end{table*}

In this section, we focus on the co-existence region when only the GD effect is present. Our analysis is applied to the three stellar models (see Tab.~\ref{T1}). For each model, the base density, $\rho_{0}$, is kept constant, as indicated in the last row of this table. The calculations begin with $\Omega = 0.68$, incrementally increasing to explore the fast wind solution. For the $\Omega$-slow solution, the calculations utilise an initial trial velocity profile with $\beta$ values ranging from 3 to 3.5. These are known to represent the characteristics of slower winds.

To compute the $\Omega$-slow solutions, the process starts at a near-critical rotational speed ($\Omega = 0.995$) and is gradually reduced. This ensures the solution follows the $\Omega$-slow wind regime, as the velocity and density structures adapt to the lower rotation rate. This iterative approach allows for identifying the transition region where the fast solution ceases to exist and the $\Omega$-slow solution emerges. Fig.~\ref{1} illustrates the mass loss and terminal velocity as a function of $\Omega$, for all three stellar models. The different behaviour of fast and $\Omega$-slow solutions is clearly seen. Moreover, co-existing regions where both types of solutions overlap are identified. As illustrated in Fig~\ref{1}  and detailed in Tab.~\ref{T2}, the co-existence interval between the fast and $\Omega$-slow solutions varies for the three stellar models, highlighting the effects of stellar parameters and rotational speed. For the B2.5 V star, the co-existence interval is observed within the range $0.936 < \Omega < 0.942$, whereas for the B0 IV star, it occurs between $0.932 < \Omega < 0.937$.\\
In contrast, for the B3 I star, the interval significantly occurs on lower $\Omega$, ranging from $0.899 < \Omega < 0.91$. The presence of GD causes the fast wind solutions to shift towards higher rotational velocities compared to the results of \citet{2018MNRAS.477..755A}, which did not account for GD effects. This shift in the fast solution also causes the co-existence interval to move towards higher rotational speeds. The inclusion of GD modifies the stellar surface flux distribution, which directly influences the effective gravity and subsequently alters the dynamics of the wind solutions. Our analysis also reveals a significant difference in the wind properties between the two solutions. Specifically, the terminal velocities of the fast solutions exceed those of the $\Omega$-slow solutions by approximately a factor of $\sim$2. 
In \citet{2018MNRAS.477..755A}, it was observed that for all three models, the mass loss rate remained nearly constant for the $\Omega$-slow solutions. However, in the presence of GD, a different trend emerges in our calculations: the mass loss rate gradually increases as the rotational speed of the $\Omega$-slow solutions increases.

This finding suggests that GD, which arises due to the influence of centrifugal forces causing the star’s equator to rotate faster than its poles, significantly affects its wind structure. As a result, the temperature distribution becomes dependent on co-latitude, leading to variations in the mass-loss rate of stellar winds. As rotation speed increases, the enhanced GD effect causes the star's surface temperature to vary more significantly between the equator and poles. This temperature variation affects the atmospheric structure and consequently impacts the mass loss rate of the star.
Therefore, the original models \citep{2018MNRAS.477..755A} without GD showed a plateau in mass loss rate for $\Omega$-slow solutions, and the inclusion of GD effects resulted in a progressive increase in mass loss rate with increasing rotational speed. 

\begin{figure}
    \centering
    \includegraphics[width=\linewidth]{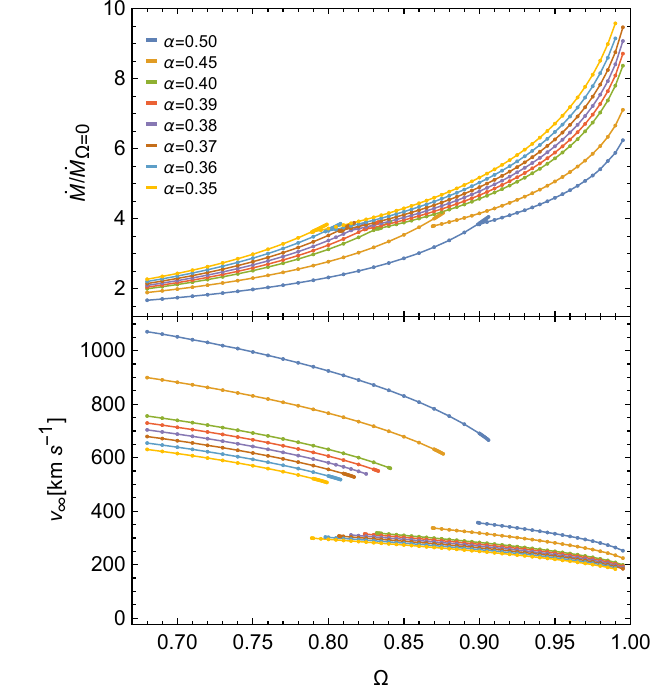}
    \caption{Line-force parameter effects on wind solutions.
    The upper panel displays mass-loss rates normalised to the non-rotating case, and the lower panel shows terminal velocities. Solutions were derived from the B0 IV model in the presence of the GD, with systematically varied line-force parameter $\alpha$ at constant $k=0.3$.}
    \label{2}
\end{figure}

\begin{figure}
    \centering
    \includegraphics[width=\linewidth]{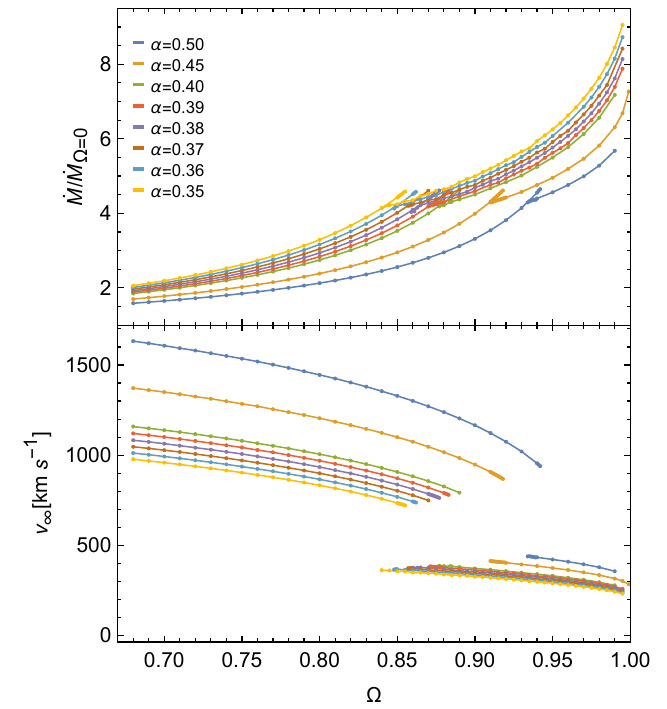}
    \caption{Line-force parameter effects on wind solutions.
    Same as Fig. \ref{2}, but for the B2.5 V model.}
    \label{3}
\end{figure}

\begin{figure}
    \centering
    \includegraphics[width=\linewidth]{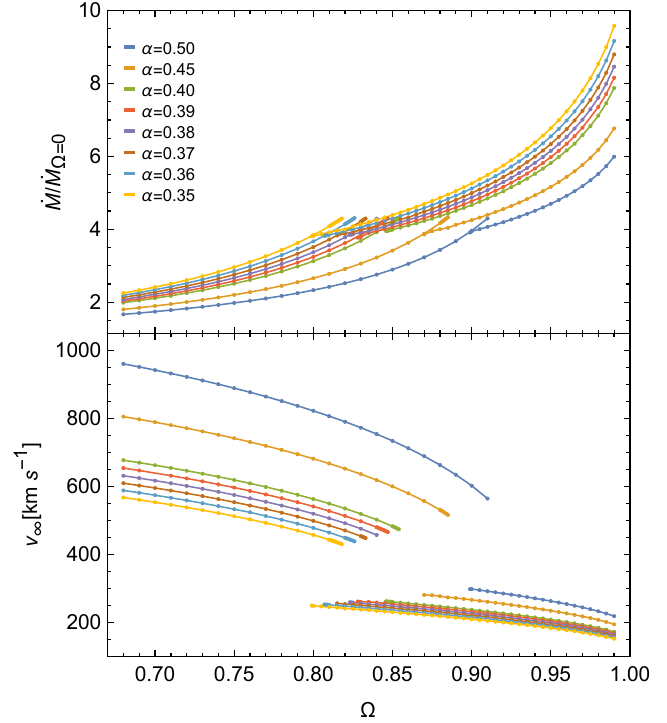}
    \caption{Line-force parameter effects on wind solutions.
    Same as Fig. \ref{2}, but for the B3 I model.}
    \label{4}
\end{figure}

Furthermore, we investigate the effect of the line-force parameter $\alpha$ on the co-existence region between the fast and $\Omega$-slow solutions. Figures~\ref{2}, \ref{3}, and \ref{4} present the mass-loss rates and terminal velocities as a function of the rotational speed $\Omega$ for all stellar models. These solutions are computed with a value of $k=0.3$ and by decreasing the value of $\alpha$ from 0.5 to 0.35. Our results show that as the value of $\alpha$ decreases, the co-existence region shifts towards the lower values of $\Omega$. This shift indicates that for the higher values of $\alpha$, higher rotational speeds are required for transition into the $\Omega$-slow solution. This behaviour is consistent with the findings of \citet[][]{2004ApJ...614..929C, 2018MNRAS.477..755A}, which state that for a weaker line force (lower $\alpha$), lower rotational speeds are sufficient to achieve the $\Omega$-slow solution. Since $\alpha$ is a line force parameter that influences the strength of the radiation-driven line force, a smaller value of $\alpha$ corresponds to a weaker line force, which requires less rotational speed to transition into the $\Omega$-slow solution. Conversely, when $\alpha$ is larger, the line force becomes stronger, requiring higher rotational speeds to reach the $\Omega$-slow regime.
This shift in the co-existence region demonstrates the sensitivity of the wind structure to the value of $\alpha$ and how this parameter controls the rotational speeds at which the transition from the fast solution to the $\Omega$-slow solution occurs. 

\subsection{Oblateness effect}

In this section, we focus on a single stellar model: a B0 IV star with the stellar and line-force parameters listed in Table~\ref{T1}. The calculations presented here consider only the effect of stellar oblateness.

\begin{figure}
    \centering
    \includegraphics[width=\linewidth]{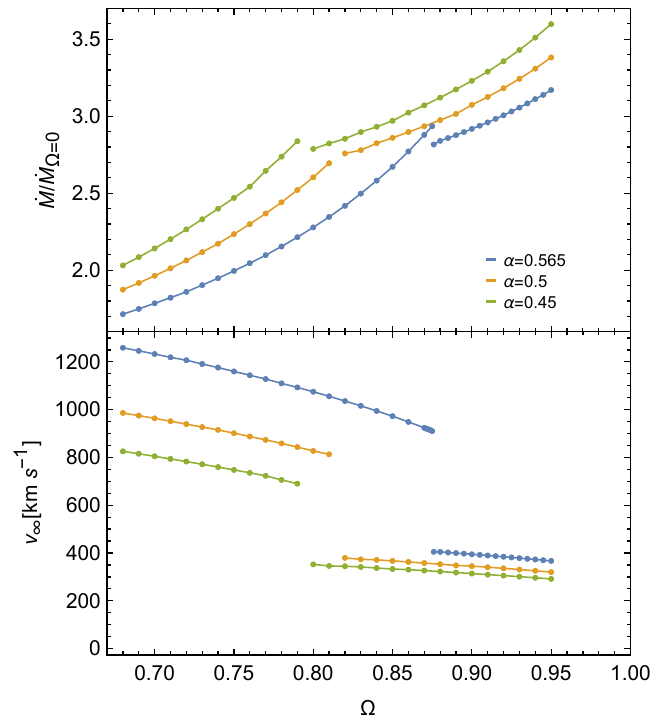}
    \caption{Stellar oblateness effect on wind parameters.
    The upper panel shows the mass-loss rates, and the lower panel shows the terminal velocities as a function of the rotation parameter $\Omega$. Calculations use the original B0 IV model parameters.}
    \label{5}
\end{figure}

As shown in Fig.~\ref{5}, the fast wind solution exists in the interval $0.68 < \Omega < 0.875$, while the $\Omega$-slow solution emerges in the range $0.876 < \Omega < 0.95$. Thus, when considering oblateness alone, no co-existence region is found between the fast and $\Omega$-slow regimes. Compared to the case with gravity darkening, the transition between the two regimes occurs at lower rotational velocities.

We further explored the influence of varying line-force parameters. Lower values of $\alpha$ shift the transition to even lower rotational speeds, and $\Omega$-slow solution emerges at even lower velocities. For instance, with $\alpha = 0.5$ and $k = 0.3$, the fast solution is found within $\Omega < 0.81$, and the $\Omega$-slow solution within $0.82 <\Omega$, as shown in Fig.~\ref{5}. Similarly, for $\alpha = 0.45$, the fast solution spans $\Omega < 0.79$, while the $\Omega$-slow solution appears in the range $0.80 < \Omega$. In both cases, no co-existence interval is observed between the two solution regimes.

\subsection{Both GD and oblate effects}

\begin{figure}
    \centering
    \includegraphics[   width=\linewidth]{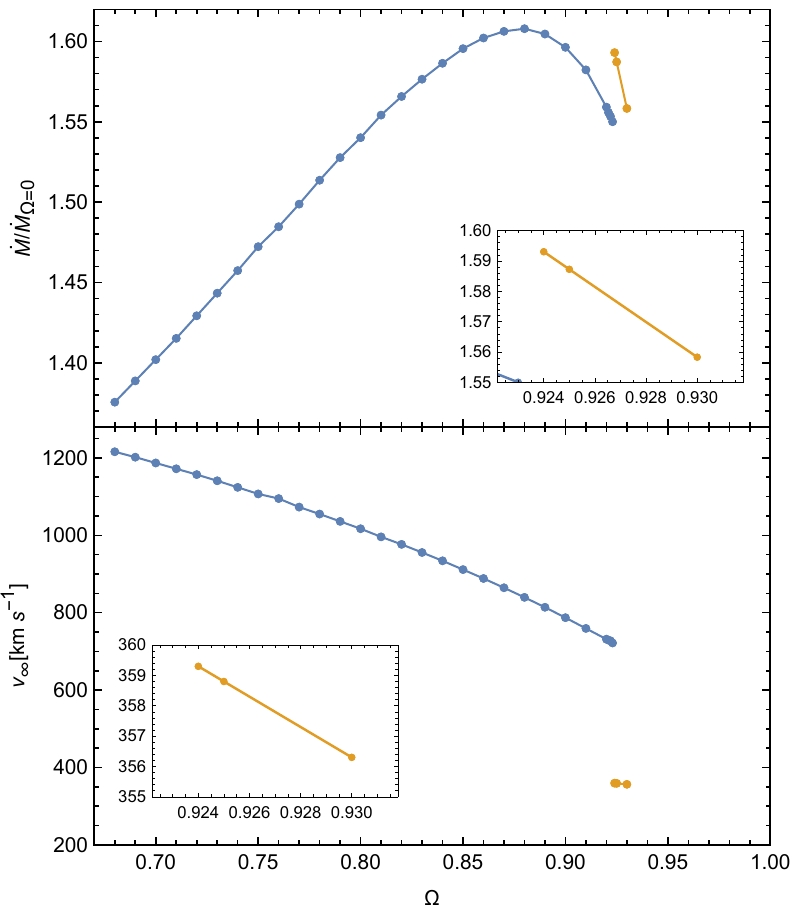}
    \caption{Combined GD and oblateness effects in the B0 IV model. The upper panel shows the mass-loss rates, and the lower panel shows the terminal velocities as a function of the rotation parameter $\Omega$. The inset panels provide a magnified view highlighting the region corresponding to the $\Omega$-slow wind solutions.}

    \label{6}
\end{figure}

In this section, we examine the combined effects of both GD and the oblate for the B0IV stellar model, using the stellar and line-force parameters provided in Tab. \ref{T1}. The fast solution is found in the interval $0 < \Omega < 0.923$,   while the $\Omega$-slow solution is present in the range of $0.924 < \Omega < 0.95$ as shown in Fig. \ref{6}. This suggests that when both GD and oblateness are considered simultaneously, the solution shifts towards higher rotational velocities compared to the oblateness effect alone, but towards lower rotational velocities than when only gravity darkening is taken into account. There is no co-existence interval in this scenario.  

Notably, the behaviour of wind parameters also differs from earlier results. When only GD or oblateness was applied, the mass loss rate increased steadily during the fast solution phase. When both effects are present, the mass loss rate initially increases during the fast solution phase, reaching a maximum, and then declines. This different behaviour is a direct result of the combined influence of GD and oblateness. When the $\Omega$-slow solution is present, the mass loss rate slightly increases compared to the fast solutions, and follows the tendency to decrease as $\Omega$ increases.We were unable to find $\Omega$-slow solutions for $\Omega$ values higher than $0.93$, possibly because we could not find a suitable initial trial solution that our numerical code required \citep{2004ApJ...614..929C}.

\section{Discussion and conclusion}
\label{sec4}

This work aims to identify the regimes in which a star’s wind transitions from the fast solution to the $\Omega$-slow solution and to determine the co-existence interval where both solutions can co-exist. 
These wind solutions are obtained by solving 1D stationary hydrodynamic equations at the stellar equatorial plane. By examining the impact of various stellar parameters, including rotational speed and line-force parameters, we aim to gain a deeper understanding of the conditions under which these distinct 1D wind solutions are present. \\

As a necessary clarification, the core goal of this work is to delineate the regimes and co-existence interval of the $\Omega$-slow solution, which appears when the rotational speed exceeds $\approx75\%$ of the critical speed \citep{2004ApJ...614..929C,2018MNRAS.477..755A,2023Galax..11...68C}. To achieve this, we utilised a 1D equatorial model, a methodological approximation that omits crucial multidimensional phenomena. Previous multi-dimensional studies, which focused primarily on the standard (fast) solutions and the Wind Compressed Disk (WCD) model \citep{1993ApJ...409..429B,1995ApJ...440..308C,1996ApJ...472L.115O}, consistently demonstrated that non-radial line-force components (NRFs) inhibit WCD formation and induce a prolate (pole-enhanced) wind morphology, often contrasting with the predictions of equatorial compression inherent to 1D models. Crucially, these findings regarding the dominance of NRFs in shaping the prolate wind structure were derived without consistently considering the specific physics (or topology) of the $\Omega$-slow solution, as the existence of this solution was established later \citep{2004ApJ...614..929C,2007ApJ...660..687M,2018MNRAS.477..755A}. Therefore, while our results map the stability and existence of the $\Omega$-slow regime in 1D, the eventual global morphology (oblate or prolate) of this denser, slower outflow remains an open question that requires future comprehensive 2D/3D modelling, including NRFs, GD and the oblate shape of the star, within the $\Omega$-slow regime. Therefore, while the 1D approximation constrains our results to the equatorial plane, it provides a physically consistent framework to isolate and characterise the onset and stability of the $\Omega$-slow regime before extending to multidimensional treatments.

According to \citet{2018MNRAS.477..755A}, the variability observed in the line profiles of massive stars can be attributed to disturbances in the photosphere, such as non-radial pulsations, stellar spots, and other perturbations. These factors, individually or in combination, can significantly alter the photospheric density when the star's rotational speed is within the co-existence range. Such density changes can potentially trigger a transition in the wind regime, particularly in the equatorial region. When the rotational velocity is within this critical range, the star may transit from a fast wind solution to an $\Omega$-slow wind solution. This transition is thought to play a key role in creating a circumstellar disc, which can form around the star's equatorial plane under these conditions. In this $\Omega$-slow wind regime, radiation pressure is the primary mechanism for mass injection into the disc. However, if the disturbances within the star's structure are not sufficiently high, the system may remain stable within the co-existence region, with the wind solution maintaining its current state. However, if the amplitude of these disturbances exceeds a certain threshold, the system might undergo another transition, reverting to the fast wind regime. This would result in the dissipation of the equatorial disc, implying that the existence and stability of these discs are closely linked to the nature of the stellar wind solution and the magnitude of the perturbations.

We analysed three different stellar models under three distinct scenarios: the presence of only GD, only the oblateness effect, and the combined effect of both GD and oblateness. Our results are compared to those obtained by \citet{2018MNRAS.477..755A}, who studied wind solutions without including these effects. We find that when GD and oblateness are included, the wind solutions exhibit a shift towards higher rotational velocities compared to the results of \citet{2018MNRAS.477..755A}. This suggests that both GD and oblateness play a substantial role in altering the wind dynamics. Furthermore, the co-existence interval, where both fast and $\Omega$-slow solutions exist simultaneously, also shifts to higher rotational velocities under these conditions. Notably, the shift in wind solutions is most pronounced where only the GD effect is considered. The combined impact of GD and oblateness also leads to significant changes, but the GD effect alone induces the most noticeable shift towards higher rotational speeds.

We also investigated the effect of varying the line-force parameter $\alpha$ on the wind solution. Our analysis reveals that the location and width of the co-existence region depend on stellar and line-force parameters as the value of $\alpha$ increases, and the wind solution shifts towards higher rotational velocities. We also find that during the $\Omega$-slow solution, the mass loss rate gradually increases in the presence of these effects. However, when accounting for oblateness and gravity darkening, the wind's mass-loss rate is lower compared to cases without gravity darkening.
As mentioned by \citet{2006ApJ...651..617A}, the mass-loss rate is primarily influenced by the equatorial mass density, which significantly affects the strength of emission profiles. For a given set of line-force parameters, increasing the rotational velocity $\Omega$ generally leads to a higher emission line intensity, whether the star is assumed to be spherical or oblate with a constant temperature. However, in the case of an oblate star with gravity darkening, the maximum line intensity is reached at an $\Omega$ value less than 1. As $\Omega$ approaches 1, the emission decreases due to the reduction in effective temperature caused by gravity darkening. Thus, the mass-loss rate decreases. 

In the context of the m-CAK theory, the presence of fast and $\Omega$ solutions, along with their co-existing interval, provides insight into stellar wind dynamics. Including GD and oblateness significantly impacts these solutions, particularly at lower line force parameter $\alpha$ values. GD reduces the radiative flux in the equatorial plane due to weaker surface gravity, while oblateness alters the stellar geometry, further weakening the driving force for winds. At lower $\alpha$, the radiative acceleration becomes insufficient to sustain outflows, especially near the equator, where the velocity drops below the threshold required for the material to escape. The absence of wind solutions for lower $\alpha$ may result from the shift in critical points within the wind equations due to physical limitations or numerical challenges in resolving the topology of the wind equations under these conditions. The terminal velocity of the $\Omega$-slow solutions is $\approx 200 $ km/s; thus, decreasing $\alpha$  leads to reduced equatorial velocities, ultimately inhibiting outflows in these regions.

In our work, we do not include the effect of viscosity in the wind equation of motion. Thus, angular momentum is conserved locally without viscosity, and the material follows purely radial outflow. In contrast, viscous forces enable the outward transfer of angular momentum, thereby reducing the terminal velocity of the wind. This is a key mechanism in the viscous decretion disk model \citep[][]{1991MNRAS.250..432L, article, 2022A&A...664A.185C}, which successfully explains the long-lived disks observed around Be stars. Thus, the inclusion of viscosity accounts for the internal angular momentum transport or the redistribution of material due to viscous forces, which can significantly influence the wind structure and disk formation. Considering this effect is beyond the scope of the present work. However, in future studies, we intend to investigate the impact of viscosity on wind solutions to gain a deeper understanding of how material flows in the equatorial region.

\begin{acknowledgements}
We thank the anonymous referee for the valuable comments and suggestions, which helped to improve the clarity and quality of this manuscript.
MC, IA, and CA also thank the ANID FONDECYT project 1230131. MC and CA acknowledge support from Centro de Astrofísica de Valparaíso - CAV, CIDI N. 21 (Universidad de Valparaíso, Chile).  AC acknowledges partial support from Centro de Estudios Atmosféricos y Cambio Climático, CIDI CR 212.115.013. AC thanks the support from Proyecto PUENTE, financiado por el proyecto PFE UVA20993, Fortalecimiento del Sistema de Investigación e Innovación de la Universidad de Valparaíso, Universidad de Valparaíso. This project was funded by the European Union (Project 101183150 - OCEANS).
Powered@NLHPC: This research was partially supported by the supercomputing infrastructure of the NLHPC (ECM-02).
\end{acknowledgements}

\bibliographystyle{aa} 
\bibliography{Ref}
\end{document}